\documentclass[12pt]{article}

\begin{document}
\title{Kinetic-theory approach to Gluon  Self-energy
  beyond Hard Thermal Loops
\thanks{This work is supported by NNSF of China under grant No. 10175026}}
\author{Zheng Xiaoping  Li Jiarong\\
        {\small {\rm a} Department of Physics, Huazhong Normal University}\\
        {\small {\rm b} Institute of Particle Physics, Huazhong Normal University}}
\maketitle
\begin{abstract}
We compare the effective dynamics of soft fields, based on temperature field theory, with the
mean field dynamics from non-Abelian kinetic theory. We derive the polarization tensor with the
leading logarithmic factor $\log({gT\over\mu})$ from the effective Boltzmann-Langevin equation
given by Litim and Manuel. The tensor is identical with effective one-loop contributions
 within the hard thermal loop effective theory.
\end{abstract}
\section{Introduction}

It is predicted that nuclear matter will transform to a quark-gluon plasma(QGP)
from hadron matter at high temperature or density.

The temperature field theory and the kinetic theory are universally used to investigate
the collective dynamics. First of all, the effective theory so-called the "hard thermal
loops"(HLT) for the soft gauge fields at the scale $gT$, where $g$ is gauge coupling and
$T$ is temperature, is constructed according to the
 fully quantum field theory techniques\cite{r1}. Soon afterwards, the effective action
 generating HTLs is straightforwardly derived from  classical kinetic theory and the
 kinetic theory  is compatible with  the  temperature
 field theory approaches in calculations of the polarization tensor\cite{r2,r3}. As well
 known,
  perturbation theory for non-Abelian gauge theories breaks down at the spatial
 momentum scale
 $g^2T$ even if the running gauge coupling $g$ is small\cite{dia1,dia2}. For static
 quantities, one can integrate out the high momentum modes($p\gg g^2T$) in perturbation
 theory using dimensional reduction\cite{dia3,dia4} and then at leading order,
 the non-perturbative
 physics associated with the scale $g^2T$ can be treated in lattice simulations,
 of which the recent results\cite{b35} show this contribution is small. However,
 dynamics quantities, such as transport coefficients, are sensitive to the scale $g^2T$
 and more difficult to deal with due to failure of dimensional reduction for
 the consideration of non-zero real frequencies. In recent years, one devoted to
 constructing effective theories of the ultrasoft gauge fields(momentum scale $\sim
 g^2T$) and fortunately obtained the effective theory by integrating out high momentum
 degrees of freedom using perturbation theory\cite{bo1}. To leading logarithmic contributions,
 the diagrammatic explanations is given\cite{bo2}. The results show that the contributions
 do not have an Abelian analogue since there are no hard thermal loop vertices in Abelian
 theories. Interestingly, there is another way to arrive at the same effective dynamics,
 that is so-called mean field dynamics from kinetic theory\cite{m1,m2}. Those furthermore
 demonstrate the connection between kinetic theory and temperature theory under specified conditions.

In this article, we will extend and apply the non-Abelian kinetic theory to compute
the self-energy of the ultra-soft modes. The polarization tensor obtained by B\"{o}deker
method is from two effective
one-loop contributions but they correspond to two 2-loop and a 3-loop diagrams in the
original theory,  which  involves  the sum over Matsubara frequencies
while the sum is not easy to compute\cite{bo2}.
 In comparison with diagrammatic approach, the kinetic-theory   appears to be much
 technically easier in calculus because some of the intrinsic complications of a
 quantum field theoretical description can be avoided\cite{m2}. Especially,
  the kinetic-theory --the temperature field theory connection can further
  be uncovered by another representative  example  beyond the "hard thermal/density loops",
  i.e. from soft field dynamics in non-perturbation domain.
  This article is organized  as follows: Sect.2 briefly reviews the the effective
   one-loop contributions to the 4-point function
   within the hard thermal loop effective theory.
   Sect.3 introduces the non-Abelian kinetic theory to compute the polarization tensor
   involving ultra-soft fields. Sect.4 gives the summary and discussion.

\section{The effective one-loop  polarization tensor}

It is well-known that the one-loop diagrams for large loop momentum($\sim T$) and soft external
momentum($\sim gT$) are called hard thermal loops(HTL's). The HTL 2-point function,
which is the same in Abelian and non-Abelian gauge theories, easily reads\cite{b16}
\begin{equation}
\delta\Pi_{\mu\nu}=m_D^2\left [-g_{\mu 0}g_{\nu 0}+p_0\int{{\rm d}\Omega_{\bf v}\over 4\pi}
{v_\mu v_\nu\over v\cdot P}\right ]
\end{equation}
where the factor $m_D$ is  Debye mass, of which the square is equal to
$(1/3)(N+N_f/2)g^2T^2$ for a SU($N$) gauge theory with $N_f$ fermions.

For momenta $p_0,p\leq gT$,  the resummatiom of the hard thermal loop can give the
correction to the tree level kinetic term. However, the investigations found that
the HTL approximation is not sufficient for obtaining the correct effective theory
for the softer modes. The some higher loop contributions should be considered.
 The two- and three-diagrams by adding soft lines($\sim gT$) in
hard momenta($\sim T$) (see fig 1 and fig 3 in \cite{bo2})
 are shown to be as large as $\delta\Pi(P)$ when the external momenta $p_0$ and $p$
 are of order $g^2T$ or smaller. Fortunately, the higher loop diagrams can be regard as two
 effective one-loop diagrams with so-called hard thermal loop vertices( fig (a) and (b)).

Thus the one-loop polarization tensors corresponding respectively to two 2-loop(with
a hard loop momentum) and
a 3-loop(with two hard loop momenta) diagrams in original theory can be computed
if ones use the HTL effective theory. B\"{o}deker derived from those after complicated treatments
\begin{equation}
\Pi^{(2)}_{\mu\nu}(P)\simeq -{i\over 4{\rm\pi}}m_D^2Ng^2Tp_0\int{{\rm d}\Omega_{\bf v}\over 4{\rm\pi}}
{v_\mu v_\nu\over (v\cdot P)^2}\left [\log\left ({gT\over\mu}\right )+{\rm finite}\right ],
\end{equation}

\begin{eqnarray}
\Pi^{(3)}_{\mu\nu}(P)&\simeq& -{i\over {\rm\pi}^2}m_D^2Ng^2Tp_0\int{{\rm d}\Omega_{{\bf v}_1}
\over 4{\rm\pi}}
{v_{1\mu} \over v_1\cdot P}
\int{{\rm d}\Omega_{{\bf v}_2}\over 4{\rm\pi}}{v_{2\nu} \over v_2\cdot P}\nonumber\\
&\ & \left [\log\left ({gT\over\mu}\right ){({\bf v}_1\cdot{\bf v}_2)^2\over\sqrt{1-({\bf v}_1\cdot{\bf v}_2)^2}}
+{\rm finite}\right ],
\end{eqnarray}
where supperscripts 2 and 3 represent respectively  the 2- and 3-loop contributions
in original theory while $\mu$ in logarithmic term is a introduced scale, which denotes
a separation of ultra-soft momenta from momenta of order $gT$, such that
\begin{equation}
g^2T\ll\mu\ll gT.
\end{equation}
Adding Eqs(2) and (3), the polarization tensor at leading logarithmic accuracy
 becomes
\begin{equation}
\Pi^{({\rm LA})}_{\mu\nu}(P)=-{i\over 4{\rm\pi}}m_D^2Ng^2T\log\left ({gT\over\mu}\right )p_0
\int{{\rm d}\Omega_{{\bf v}_1}\over 4{\rm\pi}}{v_{1\mu} \over v_1\cdot P}
\int{{\rm d}\Omega_{{\bf v}_2}\over 4{\rm\pi}}{v_{2\nu} \over v_2\cdot P}I({\bf v}_1,{\bf v}_2).
\end{equation}
This expression  is identical with ref\cite{bo2} except that a prefactor $-{1\over 4{\rm\pi}}$ exists.
It is because we here adopt Litim's notation \cite{m1,m2}
\begin{equation}
\int {{\rm d}\Omega\over 4{\rm\pi}}I({\bf v}_1,{\bf v}_2)=0,
\end{equation}
with
\begin{equation}
I({\bf v}_1,{\bf v}_2)=\delta^{(2)}({\bf v}_1-{\bf v}_2)-
{4\over{\rm\pi}}{({\bf v}_1\cdot{\bf v}_2)^2\over\sqrt{1-({\bf v}_1\cdot{\bf v}_2)^2}},
\end{equation}
while it has $\int {\rm d}\Omega I({\bf v}_1,{\bf v}_2)=0$ in ref\cite{bo2} instead.


\section{Calculus of gluon self-energy based on kinetic theory}

  The  results  obtained above are based on full field theory, which  have been done in
  detail in\cite{bo2}. As we have known, the leading kinetic equations closed to thermal
  equilibrium can derive the HTL's self-energy  as well as field theory method\cite{m1}.
  Hence we here hope to reproduce the polarization tensor beyond HTL in the framework
  of classical kinetic theory.
\subsection{Classical kinetic equations in non-Abelian plasma}

The definite trajectories described with $ x(\tau ),p(\tau ) $ and $ Q(\tau ) $
for color classical particles are govern by Wong equations \cite{wang}
\begin{equation}
\label{(1)}
  m\frac {dx^{\mu }}{d\tau }=p^{\mu }\hspace {5mm},\hspace {5mm}
  m\frac {dp^{\mu }}{d\tau }=gQ^{a}F^{\mu \nu }_{a}p_{\nu }\hspace {5mm},
     \hspace {5mm}
  m\frac {dQ^{a}}{d\tau }=-gp_{\mu }f^{abc}A^{\mu }_{b}Q_{c}
 \end{equation}

From Liouville's theorem
 $ \frac {df}{d\tau }=0 $, we can immediately get
\begin{equation}
  p^{\mu }[\partial _{\mu }-gf^{abc}A^{b}_{\mu }Q_{c}\partial ^{Q}_{a}
     -gQ_{a}F^{a}_{\mu \nu }\partial ^{\nu }_{p}]f(x,p,Q)=0,
\end{equation}
where,  $ f_{abc} $ are the
 structure constants of  SU($N$), $ A^{a}_{\mu } $ denotes the microscopic vector gauge field,
 $ F^{a}_{\mu \nu }[A]=\partial _{\mu }A^{a}_{\nu }-\partial _{\nu }
   A^{a}_{\mu }+gf_{abc}A^{b}_{\mu }A^{c}_{\nu } $ is the corresponding microscopic
   field strength  govern by Yang-Mills field equations
 \begin{equation}
 D_{\mu }F^{\mu \nu }(x)=J^{\nu }(x).
\end{equation}
It couples to the kinetic equation (9) by the currents
\begin{equation}
  J^{\mu }_{a}(x)=g\int dPdQp^{\mu }Q_{a}f(x,p,Q) .
\end{equation}
with the integral measure ${\rm d}P\equiv {\rm d}\tilde{P}{{\rm d}\Omega\over 4\pi}$,
where ${\rm d}\tilde{P}
= 4\pi{\rm d}p_0{\rm d}|{\bf p}||{\bf p}|^22\Theta(p_0)\delta(p^2)$

In principle, the set of equations can be applied to all
collective dynamics except quantum effects. When ones expand the
distribution function $f(x,p,Q)$ in powers of $g$, a polarization tensor formulated
as eq(1) can easily be derived  from the Boltzmann
equation for $f^{(1)}$\cite{r2, r3}, the distribution function at leading order in $g$.
\subsection{Beyond HTL }
The deviation $f^{(1)}$ from equilibrium distribution function is thought of as a small
fluctuation closed to equilibrium state. Therefore the excitation induced by the
fluctuation is of order $gT$. This just corresponds to the situation of HTL approximation.
However the ultra-soft field(or mean field called in\cite{m1,m2}) dynamics can be constructed
in the framework of kinetic theory if we  take statistical average over Eqs(9),(10)and (11).
This effective transport theory gives a detail description in\cite{m2}. Here we give a slightly
different simplified version in comparison with  Litim-Manuel theory. The quantities $A^a_\mu,
f, J^\mu_a$ are decomposed into mean terms and fluctuation terms
\begin{equation}
A^a_\mu=\bar A^a_\mu+a_\mu^a,  f=\bar f+\delta f, J^\mu_a=\bar J^\mu_a+\delta  J^\mu_a,
\end{equation}

Keeping only the terms including the largest contributions,
we have the equations for mean fields and fluctuations
\begin{equation}
p^\mu (\bar D_\mu-gQ_a\bar F^a_{\mu\nu}\partial^\nu_p)\bar f=\bar\xi,
\end{equation}
\begin{equation}
p^\mu\bar D_\mu\delta f=gp^\mu Q_a f^a_{\mu\nu}\partial_\nu^p\bar f+gp^\mu a_{b,\mu}f^{abc}Q_c\partial^Q_a\bar f,
\end{equation}
\begin{equation}
D^f_\mu f^{\mu\nu}_a =\delta J^a_\nu,
\end{equation}
where $\bar D\equiv D[\bar A], D^f\equiv D[a], \bar F\equiv F[\bar A], f\equiv f[a]$ and
$\xi$ reads
\begin{equation}
\xi=gp^\mu f^{abc}Q^c\partial^Q_a a^\mu_a\delta f.
\end{equation}
For convenience,  we use the Litim-manuel's notion to define a new quantity ${\cal J}$
by the following relation
\begin{equation}
J(x)=\int{{\rm d}\Omega\over 4\pi}{\cal J}(x,v).
\end{equation}
It now is well reasoned that the equations(13) and (14) transform into the ones for
 currents ${\cal J}$ and
$\delta{\cal J}$ by  the integration of Eqs(13)and (14) over d$\tilde{P}$d$Q$
\begin{equation}
v^\mu\bar D_\mu{\bar{\cal J}}^\rho+m_D^2v^\mu v^\nu\bar F_{\mu 0}=\bar\xi^\rho,
\end{equation}
\begin{equation}
(v^\mu\bar D_\mu\delta{\cal J}^\rho)_a=-m_D^2v^\rho v^\mu f_{a,\mu 0}-gf_{abc}v^\mu a_\mu^b{\bar{\cal J}}^{c,\rho},
\end{equation}
and hence we have
\begin{equation}
\xi^\rho_a=-gf_{abc}v^\mu a_\mu^b\delta{\cal J}^{c,\rho}.
\end{equation}

From these, we can reproduce the equations found in\cite{m1,m2} follows as
\begin{equation}
v^\mu D_\mu{\cal J}^\rho(x,v)+m_D^2 v^\rho v^\mu F_{\mu 0}(x) =-\gamma v^\rho\int{{\rm d}
\Omega_{{\bf v}'}\over 4\pi}I({\bf v},{{\bf v}'}){\cal J}^0(x,v')+\zeta^\rho (x,v),
\end{equation}
Here we remove the notion bar denoting the mean fields, the $\zeta^\rho$ represents the stochastic
noise, the coefficient $\gamma$ reads
\begin{equation}
\gamma={g^2\over 4\pi}NT\log\left({gT\over\mu}\right ).
\end{equation}

Combining the  equations and the Yang-Mills equations of mean fields together, Litim and Manuel
derived the color conductivity with the logarithmic effect $\log\left({gT\over\mu}\right )$.

\subsection{Self-energy beyond HTL approximation}

 We can now proceed to do calculations from eq(21) to obtain the polarization tensor beyond
 HTL approximation. We adopted the iterating method in \cite{m2} to solve the equation(21).  The
 distinct scale parameters involving (21) are well separated
 \begin{equation}
g^2T\ll\gamma\ll gT\ll T.
\end{equation}
Then we expand ${\cal J}^\mu(x, v)$ in term of ${\gamma\over v\cdot D}$
close to the scale of $gT$
\begin{equation}
{\cal J}^\mu(x, v)=\sum\limits_{n=0}^\infty {\cal J}_{(n)}^\mu(x, v),
\end{equation}
Therefore, for momenta about the Debye mass, it immediately has
\begin{equation}
v^\mu D_\mu{\cal J}_{(0)}^\rho(x,v)=-m_D^2 v^\rho v^\mu F_{\mu 0}(x) +\zeta^\rho (x,v),
\end{equation}
while for momenta below the Debye mass, the damping term of collision integral is important,
\begin{equation}
v^\mu D_\mu{\cal J}_{(n)}^\rho(x,v)=-\gamma v^\rho\int{{\rm d}
\Omega_{{\bf v}'}\over 4\pi}I({\bf v},{{\bf v}'}){\cal J}_{(n-1)}^0(x,v').
\end{equation}
The  current ${\cal J}^\rho_{(0)}$ obeying the equation (25)
coincides with the HTL current besides the noise term. In local limit approximation, the
equation becomes in momentum space
\begin{equation}
-iv\cdot K {\cal J}^\rho_{(0)}(K, v)=-m_D^2v^\rho v^\mu F_{\mu 0}(K),
\end{equation}
Taking $\rho =0$, we have
\begin{equation}
{\cal J}^0_{(0)}(K,v)=-im_D^2{v^\mu\over v\cdot K}F_{\mu 0}(K)=m_D^2\left [
-g^{0\mu}+k_0{v^\mu\over v\cdot K}\right ]A_\mu,
\end{equation}
Substituting this into the equation for $n=1$ in (26), we get in momentum space

\begin{equation}
-iv^\mu K_\mu{\cal J}_{(1)}^\rho(K,v)=-\gamma v^\rho\int{{\rm d}
\Omega_{{\bf v}'}\over 4\pi}I({\bf v},{{\bf v}'}){\cal J}_{(0)}^0(K,v').
\end{equation}
The current reads
\begin{equation}
{\cal J}_{(1)}^\rho(K,v)=-im_D^2\gamma k_0
{v^\rho\over v\cdot K}\int{{\rm d}\Omega_{{\bf v}'}\over 4\pi}{v'^\mu\over v'\cdot K}I({\bf v},{{\bf v}'})A_\mu,
\end{equation}
here we apply integral (6). The physical current can be obtained by integral over angle
\begin{equation}
J^\rho_{(1)}=\int{{\rm d}\Omega_{\bf v}\over 4\pi}{\cal J}_{(1)}^\rho(K,v)
= -im_D^2\gamma k_0\int{{\rm d}\Omega_{\bf v}\over 4\pi}
{v^\rho\over v\cdot K}\int{{\rm d}\Omega_{{\bf v}'}\over 4\pi}{v'^\mu\over v'\cdot K}I({\bf v},{{\bf v}'})A_\mu.
\end{equation}
Compared this with the definition
\begin{equation}
J^\mu=\Pi^{\mu\nu}A_\nu,
\end{equation}
a polarization tensor is arrived at
\begin{equation}
\Pi^{\mu\nu}_{(1)}(K)=-im_D^2\gamma k_0\int{{\rm d}\Omega_{\bf v}\over 4\pi}
{v^\rho\over v\cdot K}\int{{\rm d}\Omega_{{\bf v}'}\over 4\pi}{v'^\mu\over v'\cdot K}I({\bf v},{{\bf v}'}).
\end{equation}

This result evidently coincides with the $\Pi^(LA)$ formulated in (5).

Continuous iteration, the higher currents are easily express in powers of ${\gamma\over v\cdot K}$
\begin{equation}
J^\rho_{(n)}= -im_D^2 k_0
\int\left [\prod\limits^{n}_{j=1}{{\rm d}\Omega_{{\bf v}_{j}}\over 4\pi}{\gamma\over v_{j}\cdot K}I({\bf v}_{j+1},{{\bf v}_j})\right ]
v^\rho_nv^\mu_1A_\mu.
\end{equation}

\section{Summary and Discussion}

    The link between the temperature field theory and the kinetic theory for QCD plasma is one of
the probing subjects. Ones have realized that the field theory is identical with the classical
kinetic theory at level of HTL's approximation. Recently, ones found respectively effective
dynamics of soft fields or mean field dynamics from the diagrammatic approach and the kinetic
theory. However the self-energy beyond HTL is given from the field theory but not from the kinetic
theory.

    Here we firstly compare the field theory with the kinetic theory method and find the
calculations of polarization tensor in classical kinetic theory is simpler than the field theory.
Secondly, we re-analyze the mean field dynamics from the kinetic theory
based on the understanding of our simplified version. We follow the Litim-Manuel's philosophy
to expand  the effective Boltzmann equation\cite{m2} and slightly proceed to derive the polarization tensor
containing the logarithmic factor $\log{gT\over\mu}$. Our result fully coincides with the diagrammtic
approach based on the full field theory. Remarkably,  successively integrating out the hard
and semi-hard(or called soft) degrees of freedom must be adopted in both calculations to obtain the dynamics of
ultra-soft fields, but the advantage in the kinetic theory is  able to directly extract the
polarization tensor while much complicated treatments have to be done in the diagrammatic approach.


\begin{thebibliography}{aa}
\bibitem{r1} F. Abe et al., Phys. Rev. Lett. 71,3421(1993).
\bibitem{r2} P. F. Kelly ,Q. Liu ,  C. Lucchcsi, C. Manuel,
              Phys. Rev. Lett.,72, 3461(1994).
\bibitem{r3} P. F. Kelly ,Q. Liu ,  C. Lucchcsi, C. Manuel,
              Phys. Rev., D50, 4209(1994).
\bibitem{dia1}A. D. Linde, Phys. Lett. B96,289(1980).
\bibitem{dia2}D. J. Gross, R. D. Pisarski and L. G. Yaffe, Rev. Mod. Phys. 53,43(1981).
\bibitem{dia3}K. Farakos, K. Kajantie, K. Rummukainen and M. Shaposhnikov, Nucl. Phys. B425,67(1994).
\bibitem{dia4}E. Braaten, A. Nieto, Phys. Rev. D51, 6990(1995).
\bibitem{b35} K. Kajantie, M. Laine, K. Rummukainen and Y. Schroder, Phys. Rev. Lett. 86,10(2001).
\bibitem{bo1}D. B\"{o}deker, Phys. Lett. B426,351(1998).
\bibitem{bo2}D. B\"{o}deker, Nucl. Phys.  B566,402(2000).
\bibitem{m1} D. F. Litim and C. Manuel, Phys. Rev. Lett. 82,4981(1999).
\bibitem{m2} D. F. Litim and C. Manuel, Nucl. Phys. B562, 237,(1999).
\bibitem{b16} See, e.g.,H. A. Weldon, Phys. Rev. D26,1394(1982).
\bibitem{wang} S. Wang, Nuovo Cimento 65A,689(1970).
\end{thebibliography}
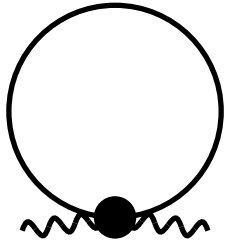
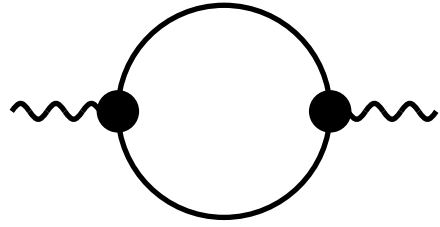
\end{document}